\documentclass{article}

\usepackage{arxiv}

\usepackage[utf8]{inputenc} 
\usepackage[T1]{fontenc}    
\usepackage{hyperref}       
\usepackage{url}            
\usepackage{booktabs}       
\usepackage{amsfonts}       
\usepackage{nicefrac}       
\usepackage{microtype}      
\usepackage{lipsum}		
\usepackage{graphicx}
\usepackage{cite}
\usepackage[version=4]{mhchem}
\graphicspath{ {./images/} }

\title{Anisotropy of acoustic properties of magnetized magnetic fluids with ellipsoidal aggregates}

\author{ \href{https://orcid.org/0000-0001-8915-2411}{\includegraphics[scale=0.06]{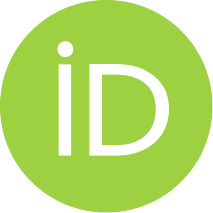}\hspace{1mm}Alexander D.~Kurilov}\\
	Federal State University of Education\\
	105005, Russia, Moscow, 10a Radio St. \\
    Prokhorov General Physics Institute\\
    of the Russian Academy of Sciences\\
    119991, Russia, Moscow, 38 Vavilova St.\\
	\texttt{ad.kurilov@gmail.com} \\
	\And
	Anastasia V.~Gubareva \\
	Federal State University of Education\\
	105005, Russia, Moscow, 10a Radio St. \\
    \And
	Sergei A.~Zubkov \\
	Federal State University of Education\\
	105005, Russia, Moscow, 10a Radio St. \\
    \And
	Denis N.~Chausov \\
	Prokhorov General Physics Institute\\
    of the Russian Academy of Sciences\\
    119991, Russia, Moscow, 38 Vavilova St.\\
}

\begin{document}
\maketitle

\begin{abstract}
A model of sound propagation in a magnetized magnetic fluid containing ellipsoidal aggregates is proposed. The model quantitatively describes the geometry of the aggregates formed from nanoparticles. Expressions for the attenuation coefficient and the sound velocity have been derived, taking into account dipole-dipole interaction between the aggregates. The model demonstrates good agreement with the experimental data. In limiting cases, the derived expressions reduce to classical ones, and the fitting parameters are merely geometric characteristics of the aggregates. The developed model enables the determination of aggregate sizes and the distances between them from an experiment that can be conducted without direct contact. The results obtained can also be used to analyze the field dependencies of acoustic properties and to model the kinetics of aggregate growth in a magnetic field.
\end{abstract}


\section{Introduction}
The propagation of elastic waves in magnetic fluids occurs through a number of physical processes. In non-magnetized magnetic fluids, the visco-inertial mechanism plays a key role~\cite{rytov1938sound,urick1948absorption,ahuja1972formulation}. This mechanism is associated with the translational motion of particles relative to the carrier fluid and is described by the viscous-inertial forces applied to the particles by the fluid: the Stokes’ drag force, the Basset force, which depends on the history of motion, and the added mass force~\cite{boussinesq1885resistance,basset1888treatise}.

From a physical point of view, the viscous-inertial mechanism arises from the difference in the densities of the dispersion medium and the dispersed phase, which causes a velocity gradient between the components of the system. As a result, viscous waves appear at the interface, tending to equalize the velocities of the dispersed phase and the dispersion medium. The magnitude of the velocity gradient depends on the size and shape of the particles, the density difference between the phases, the viscosity of the dispersion medium and the frequency of the sound wave.

The theory of anomalous absorption of sound waves in a suspension of solid particles due to the viscous-inertial mechanism was formulated for the first time by S.\,M.~Rytov, V.\,V.~Vladimirsky and M.\,D.~Galanin~\cite{rytov1938sound}. The key parameter in this theory is the ratio of the particle radius $R$ to the viscous penetration depth $\delta_\eta$
\begin{equation*}
    \sqrt{\xi}=\frac{R}{\delta_\eta}=R\sqrt{\frac{\rho_1\omega}{2\eta_1}},
\end{equation*}
where $\rho$ is the density of the medium, $\omega$ is the circular frequency of the ultrasonic wave, $\eta$ is the shear viscosity. The subscript $_1$ denotes the carrier medium.

Two modes of sound propagation in suspensions can be distinguished depending on the parameter $\xi$: the viscous and the inertial~\cite{kytomaa1995theory}. In the low-frequency region $R/\delta_\eta\ll1$, the contribution of Stokes’ drag prevails and absorption is proportional to $\varphi\omega^2R^2/\eta_1$. At $R/\delta_\eta\geq1$, the inertial effects become significant and the Basset force begins to influence the momentum transfer in the system. In the case of $R/\delta_\eta\gg1$, the boundary layer is very thin and the absorption becomes proportional to $\varphi\sqrt{\omega\eta_1}/R$. In the case of an non-magnetized magnetic fluid, the speed of sound changes by several percent. Which is nevertheless sufficient to measure the sound dispersion caused by this mechanism in the 1--100~MHz range.

Experimental confirmation of the existence of the mechanisms under consideration has been discovered by many authors both in suspensions and in emulsions~\cite{challis2005ultrasound,atkinson1992acoustic,mcclements1989scattering,dukhin1996acoustic,ballaro1980sound,ratinskaya1962sound,koltsova1973attenuation,allegra1972attenuation,rytov1938sound,jozefczak2011temperature}. Historically, sound propagation in suspensions was studied using two approaches: a hydrodynamic model based on the coupled phase equations~\cite{rytov1938sound} and an acoustic model based on the multiple scattering theory~\cite{epstein1953absorption,allegra1972attenuation}. Numerical and analytical comparison of these approaches shows that both models are strictly equivalent for dilute systems in the low-frequency region and the considered mechanisms are simulated in the same way~\cite{valier2015sound}. For the applicability of these theories for non-magnetized magnetic fluids refer to~\cite{sokolov1994viscous,gogosov1987propagation,gogosov1987ultrasound}.

Calculation of acoustic parameters in magnetized magnetic fluids is a much more complicated problem that requires taking account of the dipole-dipole interaction and the structural absorption. It is known from numerous magneto-optical experiments that internal structure of particles is formed under the influence of external magnetic field in magnetic and magnetorheological fluids~\cite{brojabasi2015effect}. The induction of an anisotropic structure leads to a dependence of the propagation velocity and the sound absorption coefficient on the angle between the wave vector and the magnetic field direction.

After the experimental discovering of the fact that chain aggregates of nanoparticles can be formed in magnetic fluids in the presence of an external magnetic field, S.~Taketomi presented them as spherical clusters arranged along lines of force. He made an attempt to explain the occurrence of anisotropy of the ultrasonic attenuation coefficient in a magnetic fluid by the rotational and translational motion of these clusters~\cite{taketomi1986anisotropy}.

Currently, the Taketomi's theory is used by many authors to describe the experimental data on the sound absorption by magnetic fluids under the influence of external magnetic field~\cite{kudelvcik2015structure,bury2020investigation,bury2020saw,hornowski2014ultrasonic,kudelvcik2016acoustic,kudelvcik2015influence}. The popularity of this theory lies in the possibility to get a various physical parameters of magnetic fluids and clusters induced in it as well as the qualitative agreement between the experimental and theoretical angular dependences of sound absorption. However, there are several arguments against the validity of this model: the absence of the added mass force, the inconsistency of limiting cases with classical expressions~\cite{sokolov2010wave}, the discrepancy between the obtained values of Leslie dissipative coefficients and thermodynamic inequalities, etc.

\begin{figure}[h]
    \centering
	\includegraphics[width=0.75\textwidth]{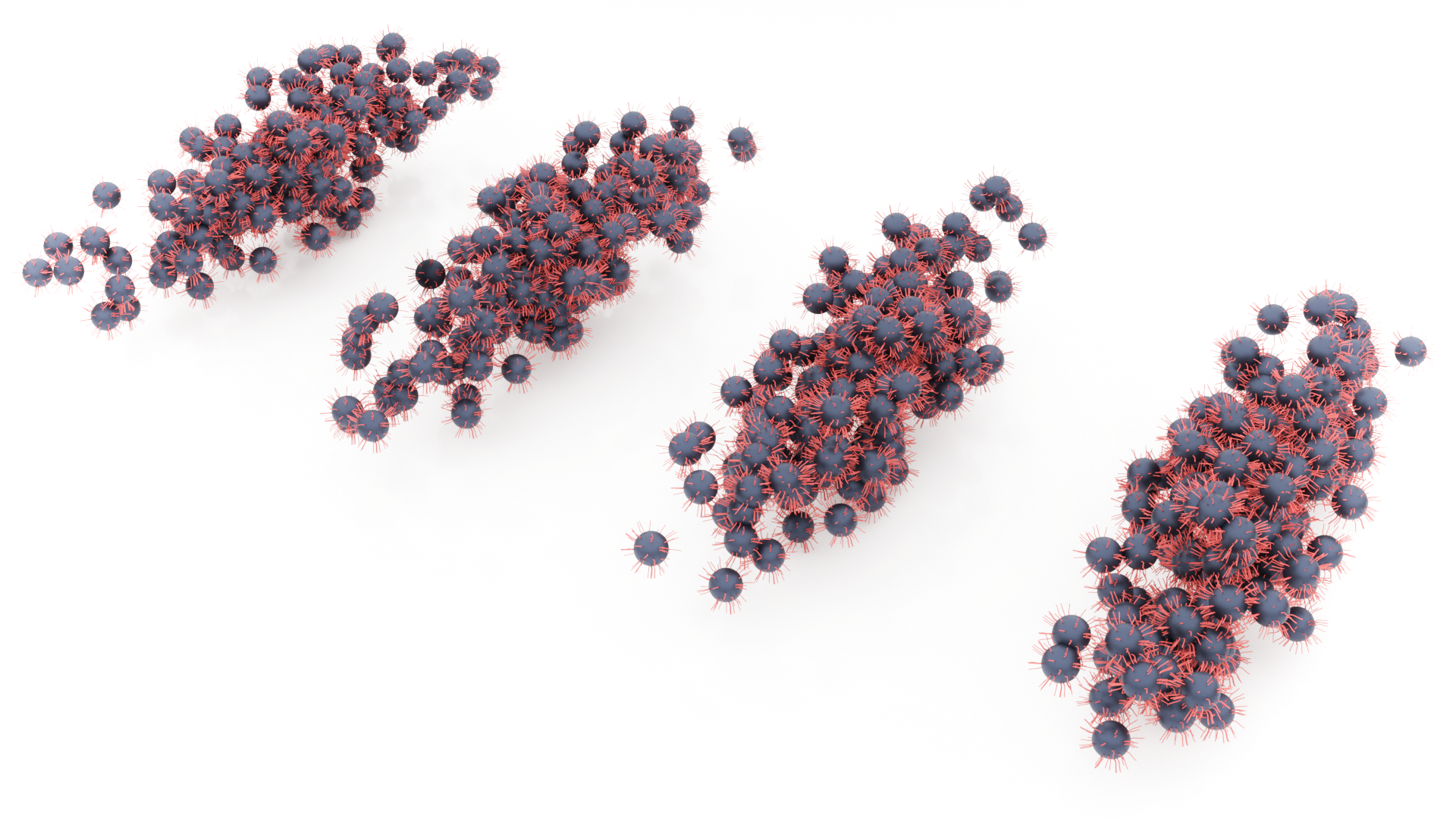}
	\caption{Scheme of an array of ellipsoidal aggregates in a magnetic fluid}
	\label{fig1}
\end{figure}

Another approach to solving this problem is to represent clusters formed in a magnetic field as a lattice of ellipsoidal aggregates. The one-dimensional case of such a lattice is shown in Figure~\ref{fig1}. A generalization of the viscous-inertial mechanism to the case of a suspension consisting of prolate or oblate spheroids was given by A.\,S.~Ahuja~\cite{ahuja1978effects}. Similar results were obtained for a magnetic fluid with ellipsoidal aggregates~\cite{gogosov1987propagation,gogosov1987ultrasound}. Afterwards, the return force of a dipole nature between the magnetized aggregates was considered, but the expression for the sound absorption coefficient was derived only for the orthogonal orientation of the wave vector and the direction of the magnetic field~\cite{kashparkova1991effect}. A further generalization to the case of an arbitrary orientation of ellipsoidal aggregates was given in~\cite{nadvoretskii1997ultrasound}.

This approach to the structural absorption appears more justified, since the aggregates induced by the magnetic field are rigid clusters elongated along the magnetic field lines.

In this work, the model for the propagation of sound in a magnetized magnetic fluid containing ellipsoidal aggregates is further developed. Based on the theory of A.S. Ahuja~\cite{ahuja1978effects}, a generalization is formulated that takes into account dipole-dipole interactions between ellipsoidal clusters in the magnetic fluid, considering their arbitrary alignment relative to the magnetic field. Expressions for both the absorption coefficient and the speed of sound propagation in the magnetic fluid are derived, which reduce to previously obtained results in limiting cases. There is satisfactory agreement between the experimental data and the results obtained according to the proposed model.

\section{Propagation of ultrasonic waves in a magnetized magnetic fluid with ellipsoidal aggregates}
For an accurate description of experimental data on sound propagation in magnetized magnetic fluids, it is necessary to have theoretical expressions for both absorption and sound velocity. This enables comparison between experimental and theoretical data on dispersion and anisotropy of acoustic parameters, as well as facilitates analysis of the involved mechanisms of sound propagation in composite media.

The introduction of return force and taking into account the ellipsoidality of the induced aggregates entails the appearance of anisotropy of both the absorption coefficient and the speed of sound in magnetized magnetic fluids. The anisotropy of the sound velocity in magnetic fluids is well known, and there are a number of theories that satisfactorily describe the experimental data~\cite{sokolov2010wave, sawada2016ultrasonic, sokolov1997anisotropy, motozawa2005influence}. The derivation of expressions for the anisotropy of the absorption coefficient and the speed of sound, as well as their simultaneous experimental verification, could shed light on which approach is correct and help formulate an actual theory of sound propagation in magnetized magnetic fluids.

\subsection{The equation of motion of an ellipsoidal aggregate}
Let us consider a lattice of ellipsoidal aggregates of the same size, aligned strictly in the direction of the magnetic field (saturation field condition).

The derivation of the generalized theory of sound propagation in magnetized magnetic fluids is based on the equation of motion of an ellipsoid in an acoustic field obtained in the manuscript by Ahuja~\cite{ahuja1978effects}
\begin{equation}
	\label{eq1}
	\rho_2\dfrac{\mathrm{d}v_2}{\mathrm{d}t}=\rho_1 \dfrac{\mathrm{d}v_1}{\mathrm{d}t}-\tau\rho_1\frac{\mathrm{d}u}{\mathrm{d}t}-s\rho_1\omega u,
\end{equation}
where the following notations are introduced
\begin{equation*}
	\tau=L+\frac{9}{4}\frac{\delta_\eta}{b}K, \quad s=\frac{9}{4}\frac{\delta_\eta}{b}K^2\left ( 1+\frac{1}{K}\frac{\delta_\eta}{a} \right ),
\end{equation*}
here $\rho$ is the density, $v$ is the velocity, $\omega$ is the cyclic frequency of the sound wave, $u=v_2-v_1$ is the relative velocity of the aggregate, $a$ is the semi-minor axis of the aggregate, $b$ is the semi-major axis of the aggregate. The index $1$ denotes the dispersion medium, while the index $2$ denotes the aggregate.

The first term on the right-hand side of equation~\eqref{eq1} describes the force due to the presence of a pressure gradient in the medium. The other terms represent the counteraction encountered by the aggregate as it moves relative to the carrier medium. It is worth noting that even if the carrier fluid is ideal ($\eta_1\rightarrow0$), the drag force in non-steady flow does not vanish. This is caused by the presence of a force proportional to acceleration, known as the added mass force~\cite{lamb1924hydrodynamics}
\begin{equation}
	\label{eq2}
	F_a=-m_i\dfrac{\mathrm{d}u}{\mathrm{d}t},
\end{equation}
where $m_i=L m_d$ is the added mass, $m_d=V_2\rho_1$ is the mass of the displaced liquid. The inertia coefficient $L$ should be calculated separately for each particle shape and its alignment relative to the sound wave. Thus, considering the added mass is necessary when analyzing the oscillatory motion of a particle in a medium. Expressions for the dynamic shape factor and the inertia coefficient for the motion of an arbitrarily aligned aggregate in the carrier medium~\cite{nadvoretskii1997ultrasound} are provided in the Appendix.

In order to formulate a correct equation of cluster motion in the case of non-stationary flow of a viscous medium, it is insufficient to consider only the Stokes drag force and the added mass force. In particular, a third term arises, which depends on the cluster's motion history and is called the Basset-Boussinesq force~\cite{lamb1924hydrodynamics}. The Basset-Boussinesq force arises due to the lag of the boundary layer when the relative velocity changes. This force is already accounted in expression~\eqref{eq1}.

The initial equation~\eqref{eq1} must be modified since it was derived for the case of a non-magnetic suspension.

The dipole moment of an ellipsoidal aggregate is determined by the expression
\begin{equation*}
	p=M_s V_2,
\end{equation*}
where $M_s$ is the saturation magnetization of the dispersed phase, and $V$ is the volume.

\begin{figure}
	\centering
    \includegraphics[width=0.75\linewidth]{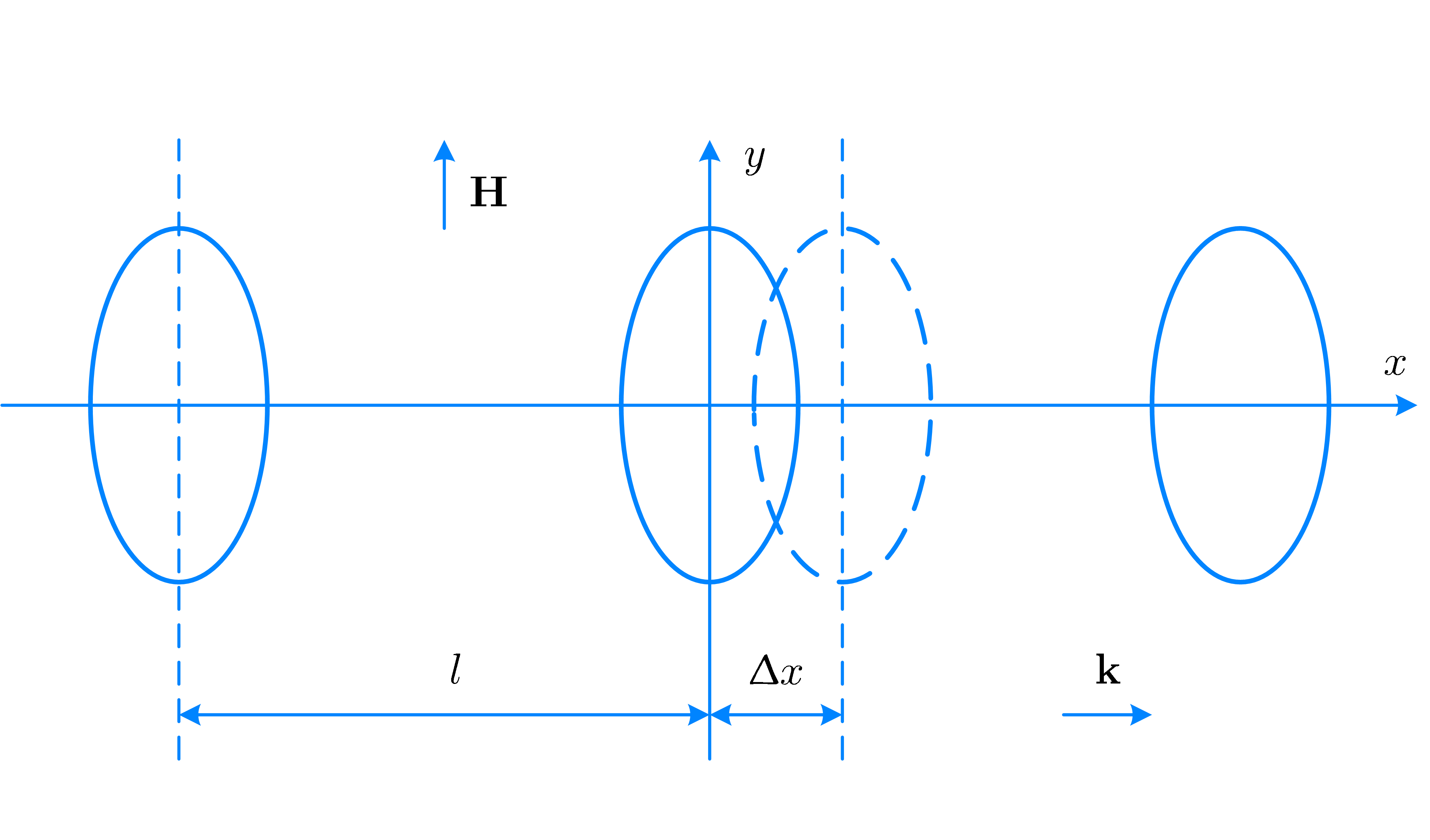}
	\caption{Scheme of sound propagation through a one-dimensional lattice of ellipsoidal aggregates}
	\label{fig2}
\end{figure}

Then, the return force of dipolar nature arising from the displacement of aggregates is determined by the expression
\begin{equation*}
	F_d=-\dfrac{6\mu_0 p^2}{\pi l^5}\Delta x\sin{\vartheta}=-\kappa\Delta x\sin{\vartheta},
\end{equation*}
where $l$ is the distance between aggregates, $\mu_0$ is the magnetic constant, $\Delta x$ is the displacement of the aggregate from the equilibrium position, $\vartheta$ is the angle between the wave vector of the sound wave and the major axis of the aggregate, and $\kappa$ is the force constant. The scheme of sound propagation through a one-dimensional lattice of ellipsoidal aggregates is presented in Figure~\ref{fig2}.

Let us rewrite the equation of motion of the aggregate taking into account the return force
\begin{equation}
	\left (\gamma+\tau \right )\dfrac{\mathrm{d}u}{\mathrm{d}t}=\left (1-\gamma \right )\dfrac{\mathrm{d}v_1}{\mathrm{d}t}-s\omega u-i\zeta\omega,
	\label{eq3}
\end{equation}
\begin{equation*}
	\zeta=\dfrac{\kappa\sin{\vartheta}}{\omega^2\rho_1 V_2}, \quad \gamma = \dfrac{\rho_2}{\rho_1}.
\end{equation*}

For sinusoidal motion $v_1=v_0 e^{-i\omega t}$, equation \eqref{eq3} can be solved for $u$ as follows
\begin{equation}
	\label{eq4}
	u=-u_0 e^{-i\left (\omega t+\psi \right)},
\end{equation}
where
\begin{equation*}
	u_0=\dfrac{\gamma-1-\zeta}{\sqrt{s^2+\left (\gamma+\tau-\zeta\right )^2}}v_0, \quad \tan{\psi}=\dfrac{s}{\gamma+\tau-\zeta}.
\end{equation*}

\subsection{Equations of mass and momentum conservation for the suspension}
Let us write the continuity equation for a suspension element, neglecting second-order terms and considering the constancy of $\rho$ in volume and time
\begin{equation}
	\label{eq5}
	-\dfrac{\partial \Delta\rho}{\partial t}=\rho \dfrac{\partial}{\partial x}\left (v_1+\varphi_2 u \right ),
\end{equation}
where $\Delta\rho=\rho\beta_S P$ is the deviation of the suspension density from its equilibrium value during the passage of an acoustic wave, with $\Delta \rho \ll \rho$, $P$ is the acoustic pressure in the suspension, $\varphi_2$ is the volume fraction of the dispersed phase, and $\beta_S$ is the adiabatic compressibility.

Substituting the expressions for $\Delta\rho$ and the solution~\eqref{eq4} for $u$, we obtain
\begin{equation}
	\label{eq6}
	-\beta_S\dfrac{\partial P}{\partial t}=\left [1-\varphi_2 \dfrac{\gamma-1-\zeta}{\sqrt{s^2+\left (\gamma+\tau-\zeta \right )^2}}e^{-i\psi} \right ]\dfrac{\partial v_1}{\partial x}.
\end{equation}

Similarly, neglecting terms beyond the first-order smallness, Let us write the equation of motion of the suspension in partial derivatives
\begin{equation}
	\label{eq7}
	\rho_1\dfrac{\partial v_1}{\partial t}+\varphi_2 \left (\rho_2\dfrac{\partial v_2}{\partial t}-\rho_1\dfrac{\partial v_1}{\partial t}\right )=-\dfrac{\partial P}{\partial x}.
\end{equation}

Replacing the substantial derivatives with partial derivatives, from equation~\eqref{eq3} we can express
\begin{equation*}
	\rho_2\dfrac{\partial v_2}{\partial t}-\rho_1\dfrac{\partial v_1}{\partial t}=-\rho_1\left (\tau\dfrac{\partial u}{\partial t}+s\omega u+i u\omega\zeta+iv_1\omega\zeta\right ).
\end{equation*}

Expanding $\partial u/\partial t$ and $u$ on the right-hand side using expression~\eqref{eq4}, we finally obtain
\begin{equation}
	\label{eq8}
	\rho_1\dfrac{\partial v_1}{\partial t}+\rho_1 \varphi_2 \left (\tau\dfrac{\partial v_1}{\partial t}+\omega s v_1\right )\dfrac{\gamma-1-\zeta}{\sqrt{s^2+\left (\gamma+\tau-\zeta\right )^2}}e^{-i\psi}=-\dfrac{\partial P}{\partial x}.
\end{equation}

\subsection{The wave equation of the suspension}
By differentiating \eqref{eq6} with respect to $x$ and \eqref{eq8} with respect to $t$, we can eliminate $\partial^2 P/\left (\partial x \partial t \right )$ and obtain the wave equation for sound propagation in the suspension:
\begin{equation}
	\label{eq9}
	\dfrac{\partial^2 v_1}{\partial x^2}=\beta_{S_1} \rho_1 \Upsilon \dfrac{1+\varphi_2 \Lambda\left (\tau+is \right )e^{-i\psi}}{1-\varphi_2 \Lambda e^{-i\psi}}\dfrac{\partial^2 v_1}{\partial t^2},
\end{equation}
where
\begin{equation*}
	\Lambda=\dfrac{\gamma-1-\zeta}{\sqrt{s^2+\left (\gamma+\tau-\zeta \right )^2}},
\end{equation*}
\begin{equation*}
	\Upsilon=1-\varphi_2 \left (1-\dfrac{\beta_{S_2}}{\beta_{S_1}}\right ).
\end{equation*}

Let us introduce the effective wavenumber $k$ of a homogeneous continuous fluid equivalent to the suspension. Then
\begin{equation*}
	k=\dfrac{\omega}{c}+i\alpha_{vd},
\end{equation*}
where $\alpha_{vd}$ is the absorption coefficient caused by viscous-inertial and dipole-dipole mechanisms, and $c$ is the speed of sound in the suspension.

It is well known that the wave equation for sound propagation in an ideal fluid can be written as
\begin{equation}
	\label{eq10}
	\dfrac{\partial^2 v_1}{\partial x^2}=\left (\dfrac{k}{\omega} \right )^2 \dfrac{\partial^2 v_1}{\partial t^2}.
\end{equation}

Comparing expressions \eqref{eq9} and \eqref{eq10}, and considering that $c_1^2=1/\left (\beta_{S_1} \rho_1\right )$, we obtain
\begin{equation}
	\label{eq11}
	\left (\dfrac{\omega}{c}+i\alpha_{vd} \right )^2=\left (\dfrac{\omega}{c_1} \right )^2 \Upsilon \dfrac{1+\varphi_2 \Lambda \left (\tau + is \right )e^{-i\psi}}{1-\varphi_2 \Lambda e^{-i\psi}}.
\end{equation}

Comparing the real and imaginary parts of \eqref{eq11} and using binomial series formulas, Euler's formula, and neglecting terms beyond the first order of smallness in concentration, we obtain the final expressions for the absorption coefficient and the speed of sound in magnetized magnetic fluids with ellipsoidal aggregates:
\begin{subequations} \label{eq12}
	\begin{equation}
		\alpha_{vd}=\dfrac{1}{2}\dfrac{\omega}{c_1}\varphi_2 \dfrac{\left (\gamma-1-\zeta \right )^2s}{\left (\gamma+\tau-\zeta\right )^2+s^2},
		\label{subeq12a}
	\end{equation}
	\begin{equation}
		\dfrac{c}{c_1}=1+\dfrac{1}{2}\varphi_2 \left ( 1-\dfrac{\beta_{S_2}}{\beta_{S_1}}\right )-\dfrac{1}{2}\varphi_2 \dfrac{\left (\gamma-1-\zeta \right )\left [\left (\gamma+\tau \right )\left (\tau +1 \right )+s^2 \right ]}{\left (\gamma+\tau-\zeta \right )^2+s^2}.
		\label{subeq12b}
	\end{equation}
\end{subequations}

By setting the densities of the aggregates and the carrier medium equal ($\gamma=1$), we obtain expressions for the absorption coefficient and the speed of sound only considering the dipole-dipole mechanism
\begin{subequations} \label{eq13}
	\begin{equation}
		\alpha_{vd}=\dfrac{1}{2}\dfrac{\omega}{c_1}\varphi_2 \dfrac{\zeta^2 s}{\left (\tau+1-\zeta\right )^2+s^2},
		\label{subeq13a}
	\end{equation}
	\begin{equation}
		\dfrac{c}{c_1}=1+\dfrac{1}{2}\varphi_2 \left ( 1-\dfrac{\beta_{S_2}}{\beta_{S_1}}\right )+\dfrac{1}{2}\varphi_2 \zeta \dfrac{\left (\tau+1 \right )^2+s^2 }{\left (\tau+1-\zeta \right )^2+s^2}.
		\label{subeq13b}
	\end{equation}
\end{subequations}

Considering the other limiting case when $\kappa=0$, the expressions in \eqref{eq12} reduce to those previously obtained in the work \cite{ahuja1978effects}. It is also noted that, within the framework of linear approximation, they fully correspond to the expressions for the viscous-inertial mechanism \cite{rytov1938sound}.

\section{Results and Discussion}
\subsection{Effect of aspect ratio and saturation magnetization of aggregates on sound propagation}

\begin{figure}
    \centering
	\includegraphics[width=0.75\linewidth]{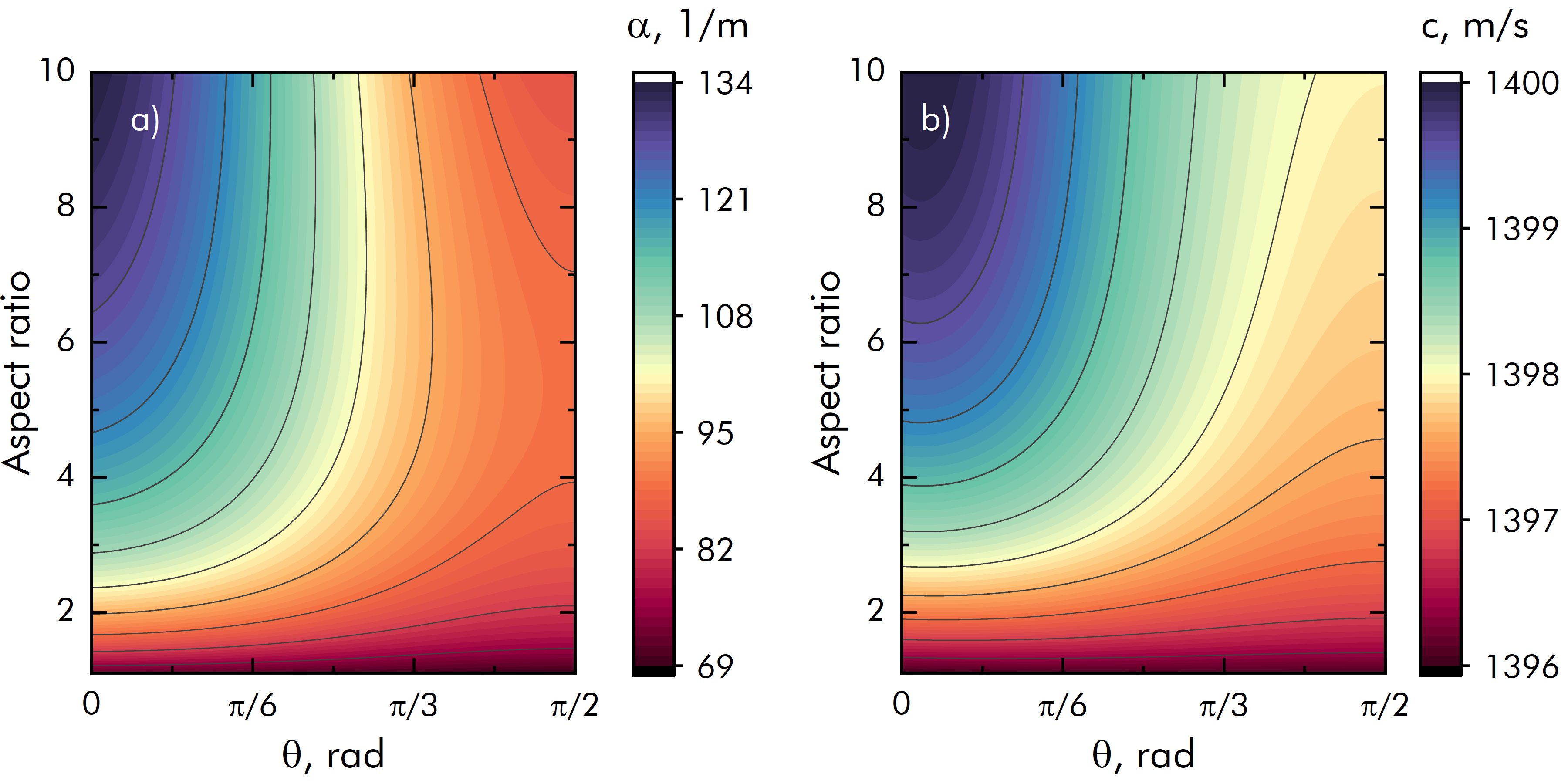}
	\caption{Angular dependencies of the absorption coefficient (a) and speed of sound (b) in a magnetic fluid with varying the aspect ratio of aggregates}
	\label{fig3}
\end{figure}

\begin{figure}
    \centering
	\includegraphics[width=0.75\linewidth]{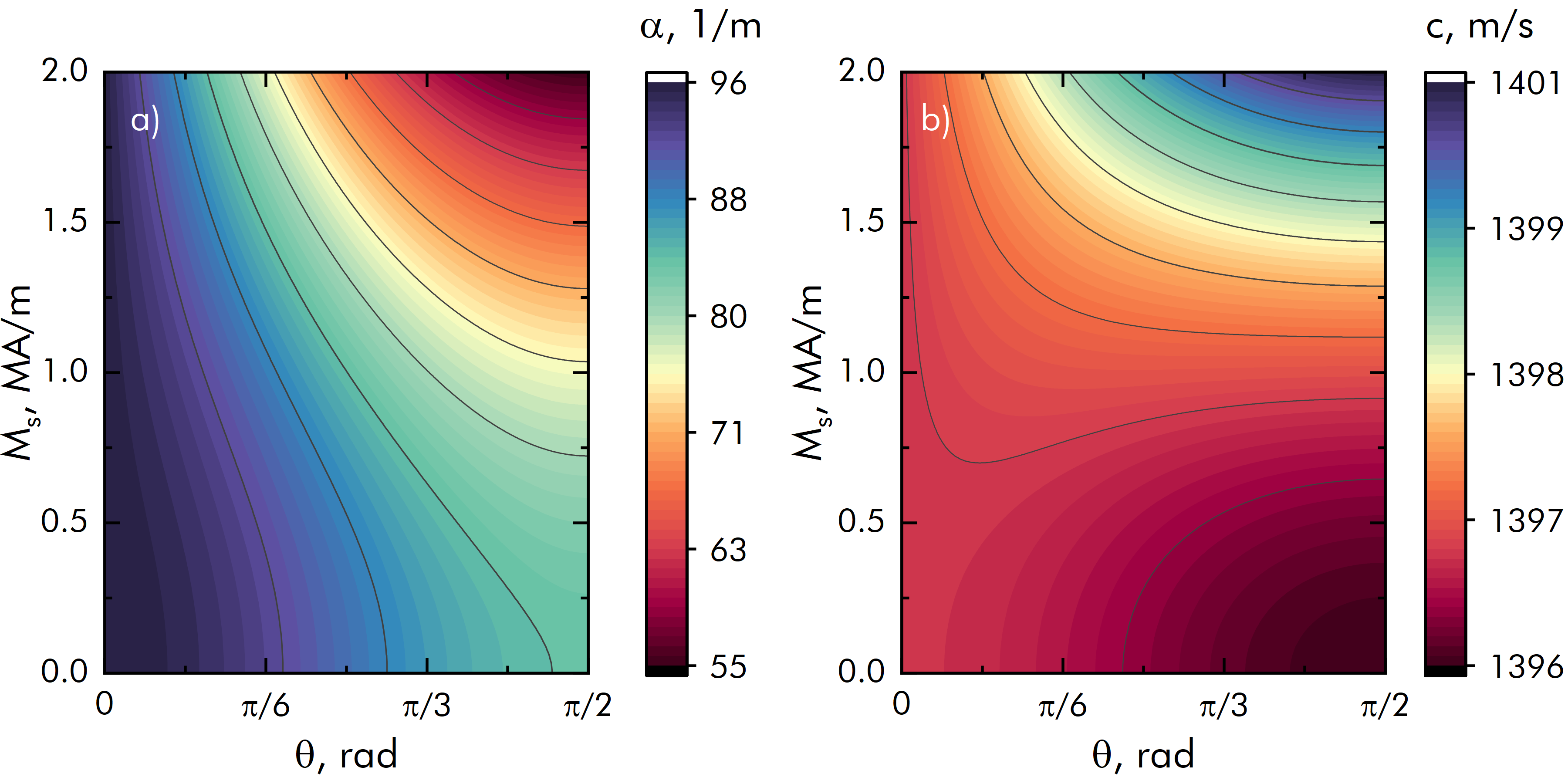}
	\caption{Angular dependencies of the absorption coefficient (a) and speed of sound (b) in a magnetic fluid with varying magnetization of aggregates}
	\label{fig4}
\end{figure}

Let us consider in more detail the derived expressions~\eqref{eq12}. The key parameters that determine the nature of the angular dependence and the magnitude of the sound absorption coefficient and sound velocity are the aspect ratio of the aggregate $h$ and its volume $V$, the density ratio between the dispersion medium and the aggregate $\sigma$, the volume fraction of aggregates $\varphi$, and the saturation magnetization of the aggregate $M_s$. The frequency of the ultrasonic wave $f$ and the shear viscosity of the carrier medium $\eta$ primarily affect the absolute values and to a lesser extent the angular dependence. To analyze the influence of these parameters on the absorption coefficient and sound velocity in detail, calculations were performed based on expressions~\eqref{eq12} by varying each parameter while keeping all others fixed. Typical parameters of a water-based magnetic fluid were used as initial values: $\rho_1=10^3$~kg/m$^3$, $\rho_2=5\cdot10^3$~kg/m$^3$, $\beta_1=5\cdot10^{-10}$~Pa$^{-1}$, $\beta_2=10^{-12}$~Pa$^{-1}$, $M_s=500$~kA/m, $\eta_1=1$~mPa$\cdot$s, $\varphi = 0.01$, $a=100$~nm, $b=200$~nm, $l=500$~nm, and $f=5$MHz. The most interesting data are those obtained by varying the aspect ratio of the aggregate $h$, as well as the saturation magnetization $M_s$, presented in Figures~\ref{fig3} and \ref{fig4}.

An increase in the aspect ratio of aggregates, on one hand, leads to an increase in the return force during the propagation of a sound wave (enhancement of the dipole-dipole mechanism), while on the other hand, it results in ambiguous changes in the influence of the viscous-inertial mechanism, which is determined by the ratio of the aggregate's semi-axes and the penetration depth of the viscous wave, as well as the alignment of the aggregates. The results in Figure~\ref{fig3} demonstrate that with the increase in aspect ratio, the anisotropy of acoustic properties significantly increases. Specifically, under collinear orientation of the magnetic field and the wave vector of the sound wave, both the absorption coefficient and the propagation velocity exhibit greater values compared to orthogonal orientation (positive anisotropy). Meanwhile, under orthogonal orientation, changes in sound velocity with increasing aspect ratio are more pronounced than changes in the absorption coefficient.

The effect of visco-inertial and dipole-dipole mechanisms on sound propagation in magnetized magnetic liquids can be shown separately in Figure ~\ref{fig4}, on which the magnetization of aggregates is varied over a wide range. An increase in magnetization leads to a significant increase in the anisotropy of the absorption coefficient and non-monotonic changes in the angular dependence of the speed of sound. Starting from a certain value of magnetization, the anisotropy of the speed of sound changes sign (with orthogonal orientation, the speed of sound becomes greater than with collinear orientation), which makes it possible to carefully isolate these mechanisms from experimental data and determine the geometric dimensions of the aggregates. It is important to note that with collinear orientation, both the absorption coefficient and the speed of sound do not depend on the magnetization of the aggregates, i.e., the dipole-dipole mechanism disappears, which can be seen analytically from expression~\eqref{eq3}.

\subsection{Comparison with experimental data}
To experimentally validate the proposed theory, we analyzed the results of our study conducted on a magnetic fluid based on transformer oil. The fluid consisted of nanoparticles with an average diameter of $16$~nm and a volume fraction of $0.2\%$. The studies were conducted at a temperature of $T=0^{\circ}$C at a frequency of $f=3.5$~MHz in a magnetic field with magnetic induction values of $0.135$~T.

\begin{figure}
    \centering
	\includegraphics[width=0.75\linewidth]{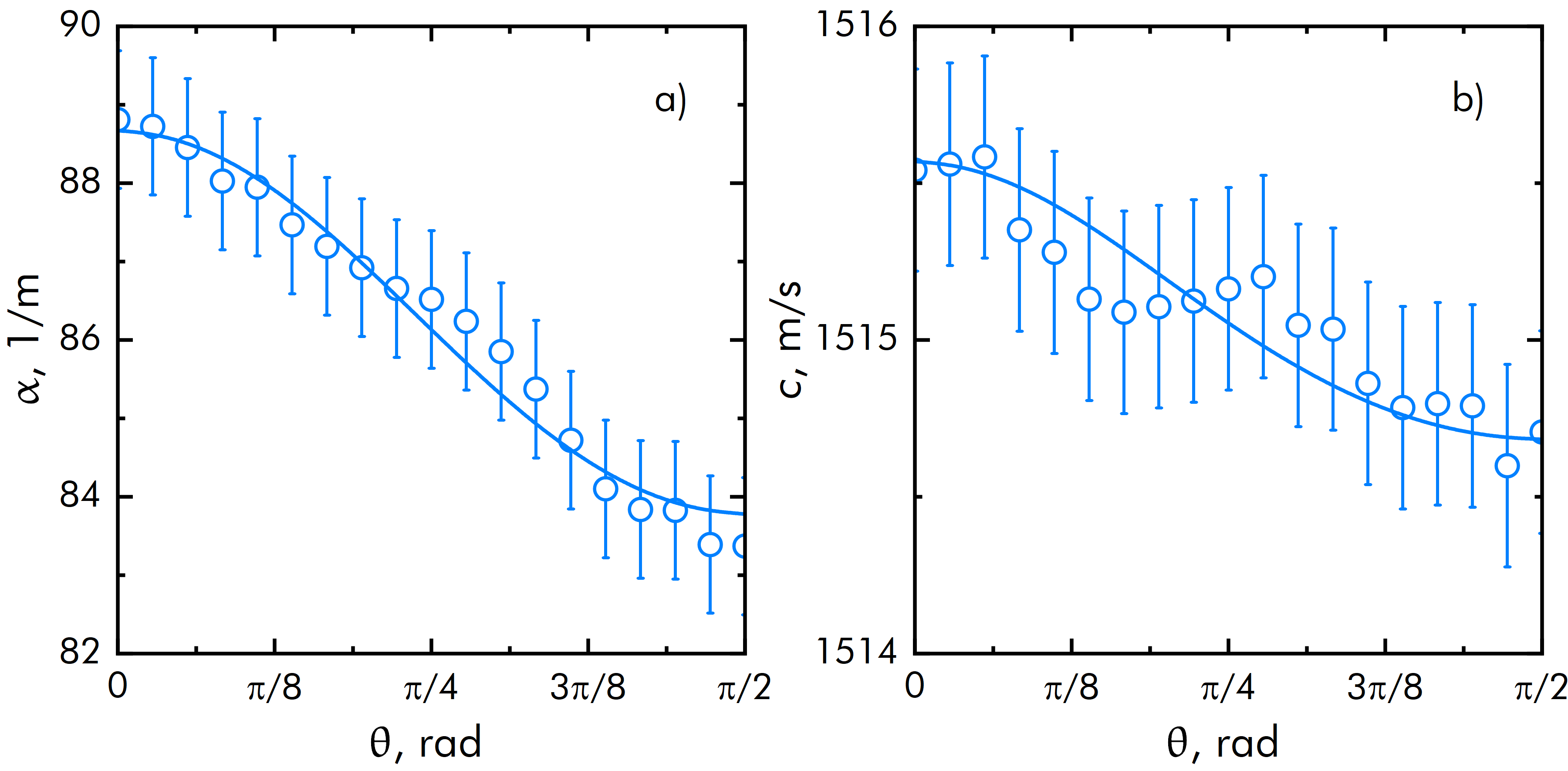}
	\caption{Angular dependence of the absorption coefficient (a) and the speed of sound (b) in a magnetic fluid based on transformer oil at $f=3.5$~MHz and $B=0.135$~T}
	\label{fig5}
\end{figure}

The angular dependencies of the absorption coefficient and the speed of sound in the magnetic fluid are shown in Figure~\ref{fig5}. The solid line represents the fitting curves corresponding to the expressions \eqref{subeq12a} and \eqref{subeq12b}. For the fitting, we used the reference values $\rho_1=864$~kg/m$^3$, $\rho_2=5050$~kg/m$^3$, $\beta_1=5.0\cdot10^{-10}$~Pa$^{-1}$, $\beta_2=10^{-12}$~Pa$^{-1}$, $M_s=480$~kA/m and $\eta_1=107$~mPa$\cdot$s. Note that the fitting parameters chosen were $a$, $b$ and $l$, which are geometric parameters of the aggregates. The volume fraction of the aggregates was not a variable parameter and was taken to be equal to the concentration of particles $\varphi_2=\varphi$.

The comparison of theoretical predictions with experimental data demonstrates strong quantitative agreement for both the absorption coefficient and the speed of sound. The values of the fitting parameters were $a=1.3~\mu$ m, $b=8.7~\mu$m and $l=17.4~\mu$m. The distances between the aggregates are comparable to their linear dimensions, and the anisotropy of the acoustic parameters is equally caused by both the anisometry of the aggregates ($b/a\approx6.6$) and the dipole-dipole interaction between them.

\section{Conclusion}
In this study, a model of sound propagation in magnetized magnetic fluids with ellipsoidal aggregates has been generalized. The anisotropy of acoustic properties in this model arises both from dipole-dipole interactions between aligned aggregates and from the orientation of ellipsoids relative to the direction of sound propagation. The proposed model does not include any effective medium parameters or numerical fitting parameters; the only variables are the geometric parameters of the aggregates.

Expressions for both the speed of sound and the attenuation coefficient in magnetized magnetic fluids are obtained. These expressions demonstrate good quantitative agreement with the experimental data. The obtained expressions can be used to analyze the field dependencies of acoustic properties and to model the kinetics of aggregate growth in a magnetic field. In addition, the observed qualitative changes in the angular dependencies of the sound absorption coefficient in varying magnetic fields can be described by the partial destruction of ellipsoidal aggregates and their structural rearrangements using the proposed model. Future research will focus on the experimental verification of the derived expressions in the case of a rotating magnetic field~\cite{sokolov2021absorption}.

The developed model enables the determination of aggregate sizes and the distance between them from a non-contact experiment. Obtaining such data is particularly important in the development of magnetofluidic microelectromechanical systems, where aggregation processes significantly affect the performance and reliability of the devices.

\section*{Acknowledgments}
This research was supported by the Russian Science Foundation grant No. 24-29-00178 (\url{https://rscf.ru/project/24-29-00178/}). We are grateful to Professor V.\,V. Sokolov for useful comments and fruitful discussions during this study.

\bibliographystyle{plain}
\bibliography{DeoLib}

\end{document}